\documentclass[12pt,preprint]{aastex}
\usepackage{natbib}

\shorttitle{DARK MATTER SELF-INTERACTION CROSS-SECTION}
\shortauthors{RANDALL \& MARKEVITCH}

\begin{document}

\title{Constraints on the Self-Interaction Cross-Section of Dark
  Matter from Numerical Simulations of the Merging Galaxy Cluster 1E~0657-56}

\author{Scott W. Randall\altaffilmark{1}, 
  Maxim Markevitch\altaffilmark{1,2}, 
  Douglas Clowe\altaffilmark{3,4}, 
  Anthony H. Gonzalez\altaffilmark{5}, 
  and Marusa Brada\v{c}\altaffilmark{6}}

\altaffiltext{1}{Harvard-Smithsonian Center for Astrophysics, 60
  Garden St., Cambridge, MA 02138, USA}
\altaffiltext{2}{Space Research Institute, Russian Academy of Science,
  Profsoyuznaya 84/32, Moscow 117997, Russia}
\altaffiltext{3}{Steward Observatory, University of Arizona, 
  933 N. Cherry Ave., Tuscon, AZ 85721, USA}
\altaffiltext{4}{Department of Physics and Astronomy, Ohio University, 
  Clippinger Lab 251B, Athens, OH 45701, USA}
\altaffiltext{5}{Department of Astronomy, University of Florida, 
  211 Bryant Space Science Center, Gainesville, FL 32611, USA}
\altaffiltext{6}{Kavli Institute for Particle Astrophysics and
  Cosmology, P.O. Box 20450, MS-29, Stanford, CA 94309, USA}

\begin{abstract}
We compare recent results from X-ray, strong lensing, weak lensing,
and optical observations with numerical simulations of the
merging galaxy cluster 1E~0657-56.  X-ray observations reveal a
bullet-like subcluster with a prominent bow shock, which gives an
estimate for the merger velocity of 4700~km~s$^{-1}$, while lensing
results show that the positions of the total mass peaks are
consistent with the centroids of the collisionless galaxies (and
inconsistent with the X-ray brightness peaks).
Previous studies, based on older observational datasets, have placed
upper limits on the self-interaction cross-section of dark matter per
unit mass, $\sigma/m$, using simplified analytic techniques.
In this work, we take advantage of new,
higher-quality observational datasets by running full N-body simulations
of 1E~0657-56 that include the effects of self-interacting dark matter,
and comparing the results with observations.  
Furthermore, the recent data allow for a new independent method of
constraining $\sigma/m$, based on the non-observation of an offset
between the bullet subcluster mass peak and galaxy centroid.
This new method places an upper limit (68\% confidence) of $\sigma/m <
1.25$~cm$^2$~g$^{-1}$.  If we make the assumption that the subcluster
and the main cluster had equal mass-to-light ratios prior to the
merger, we derive our most stringent constraint of
$\sigma/m < 0.7$~cm$^2$~g$^{-1}$, which comes from the consistency of the
subcluster's observed mass-to-light ratio with the main cluster's, and with the
universal cluster value, ruling out the possibility of a large
fraction of dark matter particles being scattered away due to
collisions.  Our limit is a slight improvement over the previous
result from analytic estimates, and rules out most of the $0.5 -
5$~cm$^2$~g$^{-1}$ range invoked to explain inconsistencies between
the standard collisionless cold dark matter model and observations.
\end{abstract}

\keywords{
dark matter ---
clusters: individual (1E0657-56) ---
methods: numerical ---
large scale structure of universe
}

\section{Introduction} \label{sec:intro}
The nature of dark matter, which accounts for the majority of the mass
in the Universe, is one of the major outstanding problems of
modern astrophysics.  Although it is often assumed that dark matter is
collisionless, there is no a~priori reason to believe that this is the
case, and it has been noted by other authors that a non-zero
self-interaction cross-section can have important astrophysical
implications (e.g., Spergel \& Steinhardt 2000).  
In particular, self-interacting dark matter (SIDM) has been invoked to
alleviate some apparent problems with the standard
cold dark matter (CDM) model, such as the non-observation of cuspy mass
profiles in galaxies (e.g., Moore 1994; Flores \& Primack 1994;
cf. Navarro et al.\ 1997; Moore et al.\ 1999b) 
and the overprediction of the number of
small sub-halos within larger systems (e.g., Klypin et al.\ 1999; Moore
et al.\ 1999a). Previous simulations and theoretical studies suggest
that a self-interaction cross-section per unit mass of $\sigma/m \sim
0.5 - 5$~cm$^2$~g$^{-1}$ is needed to explain the
observed mass profiles of galaxies (e.g., Dav\'{e} et al. 2001; Ahn
\& Shapiro 2003, though see also Ahn \& Shapiro 2005).  Earlier
studies have found stringent upper limits on $\sigma/m$, inconsistent
with the above range (e.g., Yoshida et
al.\ 2000a; Hennawi \& Ostriker 2002; Miralda-Escud\'{e} 2002, though
see also Sand et al.\ 2002).  However, in general these studies require
non-trivial assumptions or
statistical samples of clusters and full cosmological simulations.  

\begin{figure}
\plotone{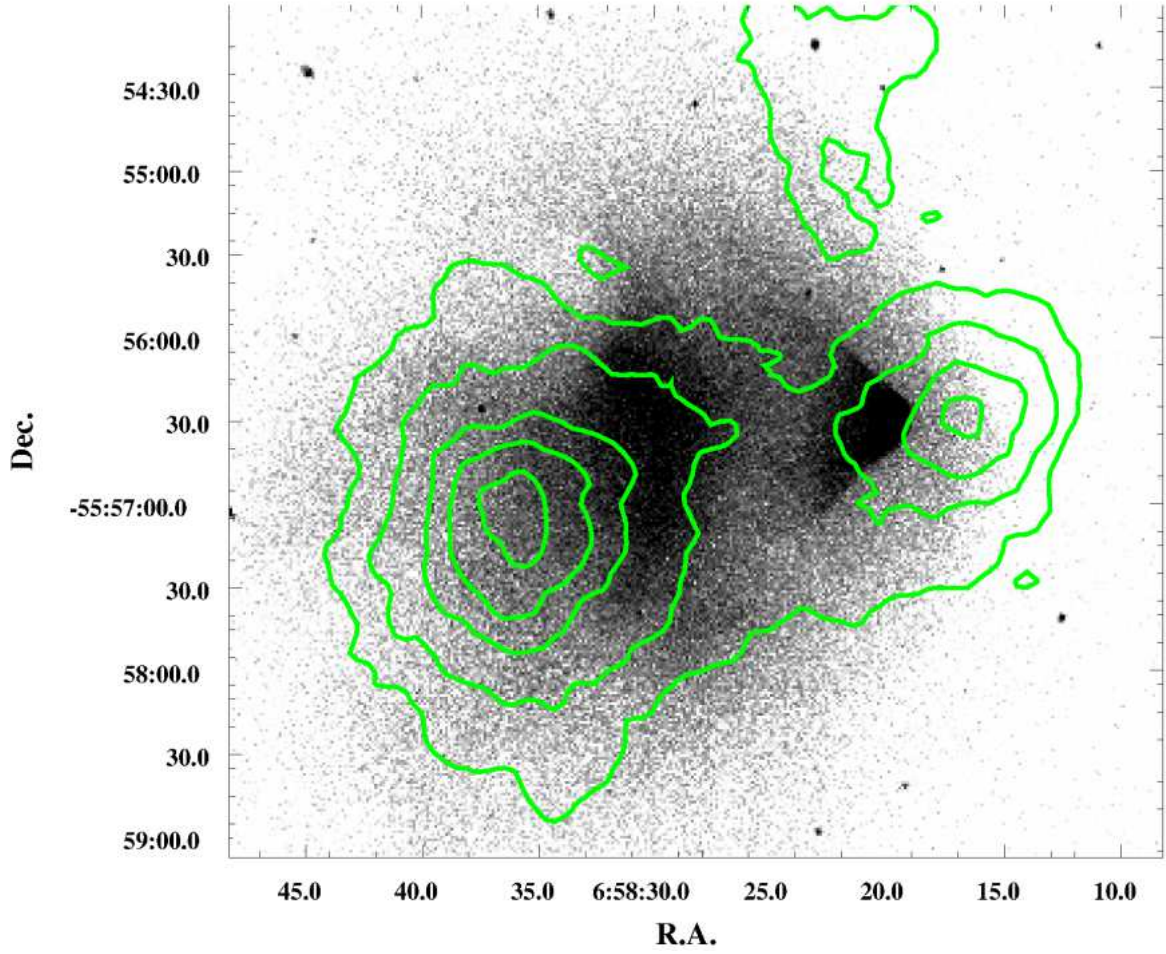}
\caption{X-ray image with weak lensing mass contours overlain.  The
  gas bullet lags the subcluster DM halo. 
  The current separation of the subcluster and main cluster mass peaks
  is 720~kpc.
\label{fig:image}}
\end{figure}

Furlanetto \& Loeb (2002) pointed out that if one observes an
offset between the gas and dark matter in a merging cluster, arising
because of the ram pressure acting on the gas but
not the dark matter, it can be used to constrain the collisional nature
of dark matter.
Markevitch et al. (2002, hereafter M02) found just such a cluster,
1E~0657-56, which in the {\it Chandra} image shows a 
bullet-like subcluster exiting the core of the main cluster, with
prominent bow shock and cold front features, and a uniquely simple
merger geometry (Markevitch et al. 2002, hereafter M02).  This gas
bullet lags behind the subcluster galaxies, which  led M02 to
suggest that this cluster could be
used to determine whether or not dark matter is collisional. If dark
matter were collisionless, one would expect the subcluster dark matter
halo to be coincident with the collisionless subcluster galaxies.
A map of the dark matter distribution was subsequently derived by Clowe
et al.\ (2004) using weak lensing observations, which
showed that the subcluster dark matter clump lay ahead of the gas
bullet, close to the centroid of the subcluster galaxies (see
also  Clowe et al.\ 2006a, hereafter C06).  The X-ray image of
1E~0657-56 is shown in 
Fig.~\ref{fig:image} with the most recently derived weak lensing mass
contours of C06 overlain.
The weak lensing contours are shown instead of the strong lensing
contours since they are better for showing the overall structure, as
they are derived from a wider field of view.
The positions of the total mass peaks from strong and weak lensing
analyses are consistent with one another, and the general
structures are similar.
The more massive main cluster is on the left and the high-velocity
merging bullet subcluster is on the right.
The main and
subcluster mass peaks are clearly visible in the mass map, as is the
offset between the gas bullet and the corresponding dark matter (DM)
peak.  C06
argued that this offset is direct evidence for the existence of dark
matter.  

The weak lensing mass map of Clowe et al.\ (2004)
was used by Markevitch et al. (2004, hereafter 
M04), in
conjunction with the X-ray and optical observations available at the time, to
analytically estimate
upper limits on the self-interaction cross-section per unit mass of
DM, $\sigma/m$, using three independent methods.
These methods were based on the
observed offset between the gas bullet and the DM subclump, the high
merger velocity of the subcluster, and the survival of the DM
subclump (more precisely, the subcluster's $M/L$ ratio being equal to
that observed in other clusters and in the main cluster).  M04
assume a King mass profile, 
based on the original weak lensing mass map, and that 
the subcluster has passed only once through the main cluster, close to
the main cluster core, as indicated by the X-ray image.   Their most
stringent limit comes from the
observed survival of the DM subclump, from which they infer that
$\sigma/m < 1$~cm$^{2}$~g$^{-1}$.

Although the analytic estimates performed by M04 provide useful
upper limits on $\sigma/m$, several conservative
simplifying assumptions were necessary.  For instance, the effects of
dynamical friction as the subcluster disturbs the main cluster mass
distribution were ignored, as was the possibility of multiple scatterings per
particle.  Although these effects are relatively small, their
inclusion may
lead to tighter constraints.  Furthermore, the analytic estimates
cannot address any structure that may be found in a high-resolution mass
map (e.g., tails in the DM distribution, similar to the gas tails seen
in the bullet, due to collisional stripping of DM, as described by
M04).  This argues for full N-body simulations that would include the
effects of SIDM with varying cross-sections.

Additionally, new data (both X-ray and lensing) have
become available for 1E~0657-56.  Analysis of data from 450 ks of
total exposure with {\it Chandra} gives a more accurate shock Mach
number of $M = 3.0 \pm 0.4$ (all uncertainties 68\%), which
corresponds to a shock
(and bullet) velocity of 4700$\pm 630$~km~s$^{-1}$ (Markevitch 2005).
Recent weak and strong lensing analyses of a much larger optical
dataset, which includes {\it HST} observations, give a 
higher quality mass 
map and a more accurate determination of the subcluster dark matter
and galaxy centroids
(Brada\v{c} et al. 2006, hereafter B06; C06).  
In particular, the accuracy of the total mass and galaxy centroids is
now sufficient for an
additional method of constraining $\sigma/m$.  In this paper, we will
concentrate on the most sensitive method from M04, which is based on
the observed mass-to-light ($M/L$) ratios, and on this new test.
The best way to interpret the new high-quality data is through
comparisons with detailed numerical simulations of the merger which
allow for SIDM with varying cross-sections.
We present results from such simulations and give
constraints on the self-interaction cross-section of dark matter
particles.  We assume $\Omega_0 = 0.3$, $\Omega_{\Lambda} = 0.7$, and $H_0
= 70$ km s$^{-1}$ Mpc$^{-1}$, for which $1\arcsec = 4.42$ kpc at the cluster
redshift of $z = 0.296$.

\section{The Simulations} \label{sec:sims}

\subsection{Simulation Code and Parameters} \label{sec:code}

All simulations were performed using a modified version of the
publicly available TreeSPH code GADGET2 (Springel 2005).  To model the
self-interaction of the DM particles, we adopted a Monte Carlo method
used previously by other authors (e.g., Burkert 2000; Yoshida et al.\ 2000b).  
At each simulation time step, the
scattering probability for the $ith$ particle is given by
\begin{equation}\label{eqn:prob}
P_i = \rho_i \sigma v_{\rm rel} \Delta t,
\end{equation}
where $\rho_i$ is the local density, $v_{\rm rel}$ is the relative
velocity between the $ith$ particle and its nearest neighbor, and
$\Delta t$ is the time step size.  The local density is determined
using GADGET2's smoothed particle hydrodynamic (SPH) capabilities.
Collisions are assumed to be elastic and scattering isotropic in the
center-of-mass frame.  In order for this relation to
be valid, $\Delta t$ must be chosen such that $P_i \ll 1$.

We ran a series of merger simulations with $\sigma/m$ varying between  0
and 1.25 cm$^2$ g$^{-1}$.  
Each simulation run included 10$^6$ DM particles 
(gas was not included in the
simulations, see discussion in \S~\ref{sec:gas}).  Additionally, we
performed a convergence test run with
10$^7$ DM particles and $\sigma/m\approx 1$ cm$^2$ g$^{-1}$, which agreed
well with the lower resolution run for all tests we performed.
We interpret this agreement as indicating
that the effects of individual self-interacting DM particles are well
modeled by the large computation particles used in the simulations,
and that the results we present here are not seriously affected by
numerical resolution effects.
The ratio
of DM particles in the main cluster and subcluster was set equal to the
initial total mass ratio of the clusters, which is known analytically
from the King models used to build the clusters.  

In this work, we apply a new method for constraining $\sigma/m$ based on
the absence of an offset between the subcluster total mass and galaxy
centroids. For this, we added another family of particles to the
simulations to represent the collisionless galaxies.
We choose 10$^5$ ``normal''
galaxy particles for the main cluster and 2.5$\times 10^4$ for the
subcluster throughout.  The ratio of the number of normal galaxy
particles was estimated based on galaxy counts given by Barrena et
al.\ (2002).  The galaxies were initially distributed like the DM in
each run.  The mass in galaxies was assumed to be roughly 5\% of the
total mass for each cluster, which, combined with the number of galaxy
particles, gives a low mass per galaxy (2.15$\times 10^8$~M$_{\odot}$).
Using a large number of light-weight galaxies was chosen over using a
more realistic mass per galaxy so that accurate galaxy centroids could
be determined.  Test simulations run with the more realistic average mass per
galaxy of $10^{11}$ M$_{\odot}$, also determined from results given by
Barrena et al.\ (2002), showed similar results to those given below in
\S\ref{sec:results}, though, as expected, with a larger scatter.  
Two cD galaxies, one at the center of each cluster, were also included
in the simulations (though we note that three cD galaxies are
observed, two associated with the main cluster and one with the
subcluster).  Their inclusion leads to
conservative estimates of the
effects of DM self-interaction, since a lower central DM density is
required to reproduce the observed total mass profile (and the
scattering probability depends on the local DM density via
Eqn.~\ref{eqn:prob}).
The cD galaxies
were each given a mass of 10$^{13}$M$_{\odot}$. 

The gravitational softening length was chosen to be 2 kpc throughout,
which is on the order of the mean inter-particle separation in the
densest region in each simulation (i.e., at core passage).  The
softening length for the cD galaxy particles was set to 60 kpc
throughout.  This large softening length was chosen since on the scale
of the simulations cD galaxies are significantly extended objects and
treating them as concentrated point-like masses would be unrealistic.
Since the lensing observations do not give an accurate mass profile
for the subcluster, King models with density profiles of $\rho(r) =
\rho_0 (1 + r^{2}/r^{2}_{c})^{-3/2}$, where 
$\rho_0$ is the central density and $r_{c}$ is the core radius, were
conservatively 
chosen for the mass profiles of each cluster.  Such a choice gives
conservative limits for 
the effects of self-interacting DM, since King models do not have strongly
concentrated ``cuspy'' cores, as compared to NFW (Navarro et al.\ 1995)
and Hernquist (1990)
profiles.  Thus the central density is lower and
the total number of DM particle collisions is conservatively reduced.  Further
discussion of the impact of the bullet mass profile on our results
can be found in \S\ref{sec:massprof}.
As suggested by the X-ray morphology (M04), all simulated mergers were
head-on collisions with zero impact
parameter and an initial separation of 4 Mpc.  
The effects of a possible non-zero impact parameter are discussed
in \S\ref{sec:impact}.

\subsection{Initial Conditions} \label{sec:ics}

For each run, the initial conditions were
chosen such that the projected mass profiles of the main cluster and
the bullet subcluster, after core passage and at the
observed separation (720 kpc), roughly matched those
from the most recent combined strong and weak lensing results derived by B06,
which are given in the last row of Table~\ref{tab:obscond}.  
A relatively small contribution from the observed distribution of gas
mass was subtracted from the B06 total mass, so 
that the resulting values could be directly compared with the
simulations.  
The gas masses were computed from the X-ray observations.
The details of the derivation of the gas mass map will be given in a
future paper.
A summary of the
parameters for each simulation run is given in Table~\ref{tab:ics},
which gives the initial central density and core radius for the main cluster
($\rho_{c,1}$, $r_{c,1}$) and the bullet subcluster ($\rho_{c,2}$,
$r_{c,2}$), and $\sigma/m$ for each run.  The mass
profiles were truncated at 20~$r_c$.

As  pointed out by C06, weak lensing is expected to underestimate the mass
of the lens by 10-20\% in the dense central regions.  Furthermore,
weak lensing can underestimate masses due to mass-sheet
degeneracy, where the mass map is affected
by the non-detection of mass at the edges of the field of view.
The effect can be seen by
comparing the total (i.e., without the gas mass subtracted) weak
lensing mass estimates of C06 to the 
mass profiles derived by B06, which combine strong and
weak lensing observations.
We chose to match the projected mass profiles in the simulations
to those given by B06, since strong lensing is expected to give better
results in the central core regions.  We are most interested in these
regions since most of the particle scattering depth accumulates near
the center.
We explore the effects of decreasing the
total mass of the system on our results in \S~\ref{sec:lowmass}.

\subsection{Matching Simulations to Observations} \label{sec:match}

Columns 4~\&~5 of Table~\ref{tab:obscond} give the total projected mass
within 150~kpc of the mass peak for the main cluster ($M_1(r < 150 $kpc$)$) and the
subcluster ($M_2(r < 150 $kpc$)$) at the 
observed separation for each simulation run and from the lensing observations.
Some trial-and-error
was necessary to determine what initial conditions gave the desired
projected mass profiles at the desired separation.  
In general, larger
$\sigma/m$ values required a more concentrated subcluster initially.
This effect occurs because the
self-interaction of the DM particles causes them to be scattered away,
particularly in high density regions.  Consequently, the subcluster
mass profile is spread out during core passage.  For the purposes of
comparing the 
simulations with observations, we take the simulation snapshot
where the offset between the clusters is closest to the observed
separation.  The time resolution of the
snapshots was small enough to match this value to within a few
kpc ($<$ 1\arcsec), which is within the observational error.
We require the simulated mass profiles at this moment to be consistent
with the strong 
lensing mass map to within 10\% in the inner regions ($r <$~500~kpc).

\subsection{Stability of Simulated Halos} \label{sec:stable}

\begin{figure}
\plotone{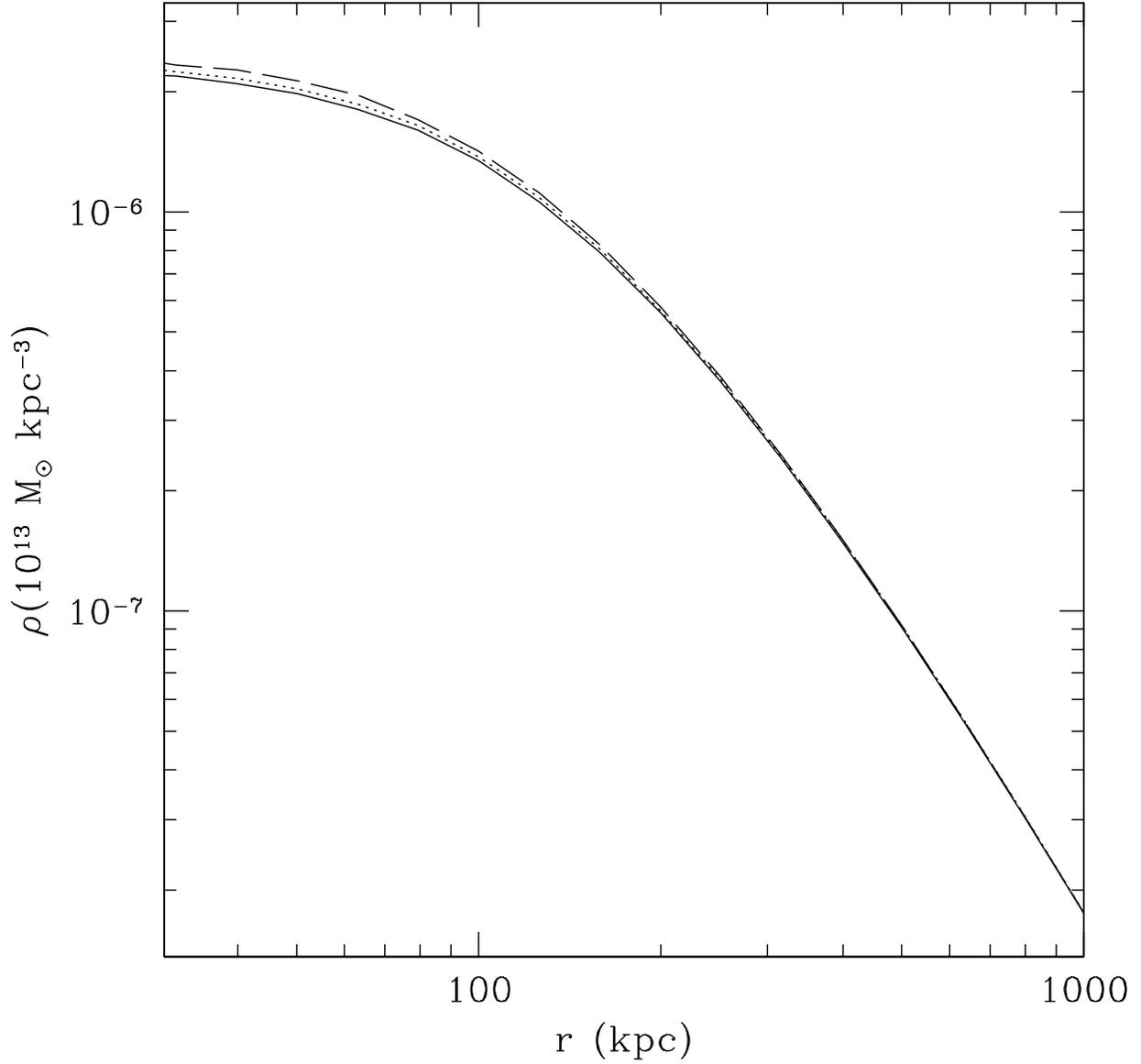}
\caption{Density profile of an isolated King model cluster at $t=0$
  (solid line), and after evolving for 1~Gyr with $\sigma/m =0$
  (dotted line) and $\sigma/m = 0.7$~cm$^2$~g$^{-1}$ (dashed line).\label{fig:rhor}}
\end{figure}

\begin{figure}
\plotone{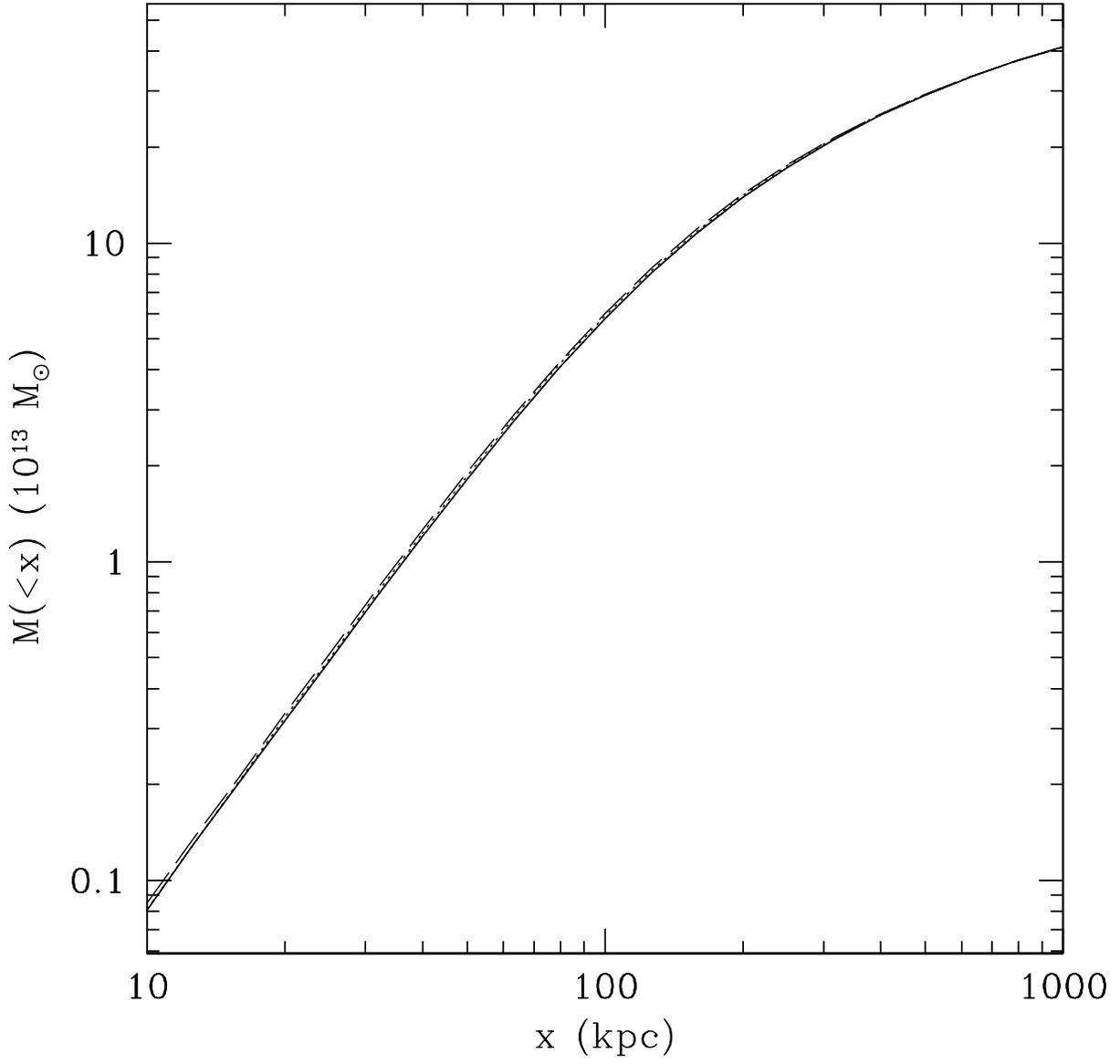}
\caption{Total mass within projected radius $x$ for the cluster
  plotted in Figure~\ref{fig:rhor}.  Line-type indications are the
  same as in Figure~\ref{fig:rhor}.\label{fig:projr}}
\end{figure}

It is of interest to evaluate the stability of our simulated clusters,
particularly in the presence of SIDM.  In the case of
non-self-interacting DM, the phase-space distribution function
can be computed and used to generate a gravitationally stable King
model density profile.  However, in the case of SIDM, particle
collisions will tend to transfer kinetic energy from one region of the
cluster to another, consequently altering the density profile (see, e.g.,
Burkert 2000).  In section \S~\ref{sec:mtol}, we will draw conclusions
based on the fraction of particles scattered away from the core of the
subcluster due to the merger event.  It is therefore necessary to
determine what fraction of particles might flow from the central
region of the bullet due to the instability resulting from SIDM
collisions.  To this
end, we ran simulations of the subcluster DM halo, allowing it to
evolve in isolation over the timescale of the merger simulations
(about 1~Gyr).  We used the same cluster parameters as the
subcluster that has the highest central density of all the
clusters in Table~\ref{tab:ics}.  The results are shown in
Figure~\ref{fig:rhor}, which gives the initial density profile (solid
line) and the density profile after 1~Gyr for $\sigma/m =
0$~cm$^2$~g$^{-1}$ (dotted line) and $\sigma/m = 0.7$~cm$^2$~g$^{-1}$
(dashed line).  The density in the inner regions is marginally enhanced
in the case of SIDM.  This result is similar to the core-collapse
phase seen by Burkert (2000), where weak interactions between the
kinematically hot core and the cooler outer regions result in an
outward transport of kinetic energy (though this effect is expected to
be somewhat curtailed here due to the near isothermality of the King
profile at small radii).  For the purposes of the test
described in \S~\ref{sec:mtol}, we are only concerned with the total
mass within {\it projected} radii.  This quantity is plotted for each
run in Figure~\ref{fig:projr}.  
The above effect of SIDM on the projected mass profile is negligible,
particularly for projected radii $x \ge 150$~kpc, which is the minimum
radius considered for the test described in \S~\ref{sec:mtol}.
Thus, if we find that a large fraction of SIDM particles scattered outside
this radius, it can be assumed to be caused by the merger event as opposed
to any halo instability.  Furthermore, the collisionless galaxies are
expected to adjust to any change in the overall potential (which is
dominated by the DM), thereby acting to further stabilize the
mass-to-light ratio.  
Indeed, in these isolated subcluster runs, the mass-to-light ratios
within a projected radius of 150~kpc from the cluster centers stay
within 2\% of their initial values, regardless of the DM self-interaction.

\section{Results} \label{sec:results}

\subsection{Galaxy -- Dark Matter Centroid Offset} \label{sec:noffset}

For non-self-interacting DM, the centroids of the subclump DM and
galaxy distributions are expected to be coincident throughout the simulation,
since gravity is the only
operating force.  However, when
$\sigma/m > 0$, the subcluster DM halo experiences a drag force as it
passes through the main cluster, and subsequently lags the
collisionless galaxies, just as the fluid-like subcluster gas core
is observed to lag the DM halo (see Fig.~\ref{fig:image}).
We ran simulations with a range of values for $\sigma/m$ and
calculated the centroids for each particle type by taking the
average projected position of the particles in some large region,
centering on this position with a smaller region, and repeating with
smaller and smaller regions (down to a region with a radius of 200
kpc).
Column~6 of Table~\ref{tab:obscond} gives $\Delta x$, the offset
between the subcluster galaxy and DM
centroids, for each run, for the moment when the subcluster is close
to the observed separation of 720~kpc from the main cluster.  
The dependence of $\Delta x$ on $\sigma/m$ is also plotted in
Figure~\ref{fig:sigma} (solid line).
Results from the run with $\sigma/m = 0$ indicate that the offsets
from the simulations are accurate to about $\pm 2$~kpc ($0.5\arcsec$).
It is clear from
Table~\ref{tab:obscond} that the centroid offset is a strong function of
$\sigma/m$.
  
\begin{figure}
\plotone{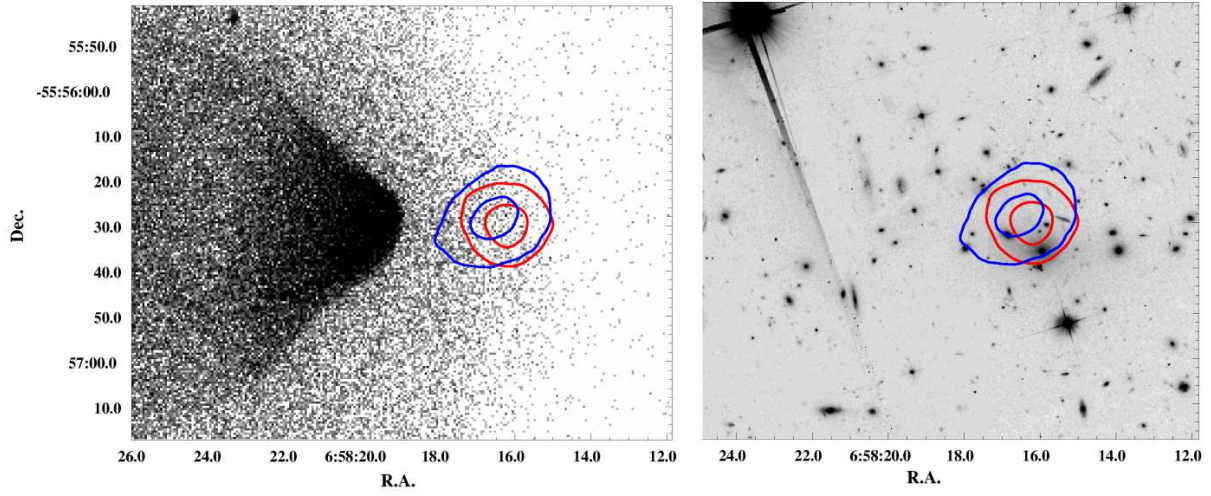}
\caption{Close up of the subcluster bullet region, with the DM (blue)
  and galaxy (red) centroid error contours overlain.
  The contours show the 68.3\% and 99.7\% error regions.  The
  left panel shows the X-ray {\it Chandra} image, while the right
  shows the optical {\it HST} image.\label{fig:bullet}}
\end{figure}

An X-ray image close-up of the bullet region with error contours
for the subcluster total mass and galaxy centroids overlain is shown
in Figure~\ref{fig:bullet}.  Details of the derivation of the total
mass centroids are given in C06.  
The centroid of the galaxy distribution was calculated from the ACS
photometry, using all galaxies for which the F814W-F606W color is within
0.15 mag of the red sequence. We used an Epanechnikov kernel with
$h=30\arcsec$ (Merritt \& Tremblay, 1994; Gonzalez et al.\ 2002) to
  determine the
centroid, and a bootstrap technique to quantify the uncertainty.
The centroid of the subcluster galaxies is found to be $5.7\arcsec \pm
6.6\arcsec$ ($25\pm29$ kpc) west of the corresponding weak lensing
mass peak.  Given
the observational errors on the centroid positions (roughly 5$\arcsec$, or
22~kpc, on the subcluster mass peak and galaxy centroid), the absence
of a larger offset means that $\sigma/m<$~1.25~cm$^2$~g$^{-1}$.  We
note that, although this upper limit is greater than the best
constraint of 
$\sigma/m<$~1~cm$^2$~g$^{-1}$ found by M04, it is more robust, since it does
not rely on the assumption that the subcluster and the main cluster
had equal $M/L$ ratios prior to the merger, as is the case with the limit
from M04 (see \S~\ref{sec:mtol}).  This distinction is relevant since,
although there is evidence for a universal $M/L$ ratio for clusters,
the level scatter for individual clusters is not negligible (see Dahle
2000).

\subsection{Subcluster M/L Ratio} \label{sec:mtol}

\begin{figure}
\plotone{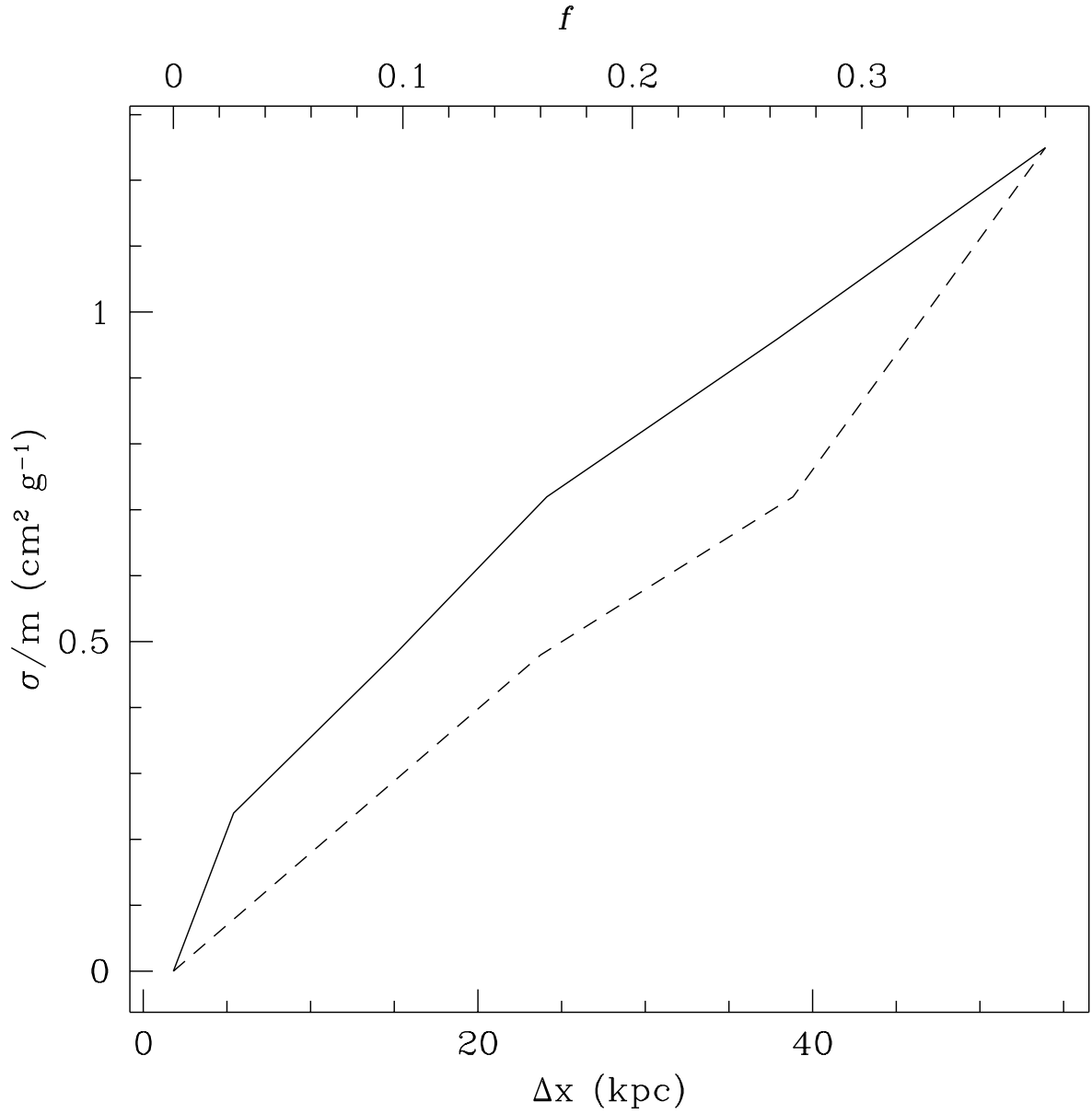}
\caption{
  The dependence of the subcluster galaxy and total mass centroid offset
  ($\Delta$ x, solid line) and the fractional change in the subcluster
  $M/L$ ratio ($f$, dashed line) on $\sigma/m$.  Based on the values
  given in Table~\ref{tab:obscond}.\label{fig:sigma}}
\end{figure}

In a merger scenario, SIDM is expected
to give a lower $M/L$ ratio for the subcluster that has just passed
through a dense core as compared to
collisionless DM.  This is because during the merger, DM particles are
scattered away due to 
collisions, while the collisionless galaxies are relatively unaffected.
To estimate the change in the $M/L$ ratio in the simulations due to
the merger, we
simply take the ratio of the total mass to galaxy mass within 150 kpc
(projected) of
the bullet DM centroid and compare the values at the start of the
simulations and at the observed separation.  The results are tabulated
in Column~7 of Table~\ref{tab:obscond}, which gives $f$, the
fractional decrease in the bullet $M/L$ ratio within 150~kpc, and also
plotted in Figure~\ref{fig:sigma} (dashed line).  We note
that for $\sigma/m \approx 1$~cm$^2$~g$^{-1}$, 
the subcluster loses about 38\% of its mass within 150~kpc, 
which is in agreement with a conservative estimate of 20~-~30\% given
by M04.  As
expected, the numerical results yield somewhat tighter constraints
on $\sigma/m$ as compared to the analytic estimates when using the same
method and observational constraints.

Using the latest lensing mass map from B06, we rederived $M/L$
ratios for each of the two subcluster within a projected 150~kpc of
the total mass peaks 
(for previous results see Clowe
et al., 2004).  For the subcluster, the mass contribution from the
outskirts of the main cluster has been approximately subtracted,
whereas for the 
main cluster, the total mass is used, since the contribution from the
subcluster is negligible.  The projected mass contribution from the
main cluster to the subcluster is estimated by taking the average mass
in an annulus at the distance of the subcluster (excluding the
region of the subcluster itself).
This gives a conservative estimate for the upper limit on $\sigma/m$,
since scattering due to putative DM collisions is expected
to result in an anomalously low $M/L$ value for the subcluster as
compared to the main cluster, and by reducing the observed mass of the
subcluster we minimize the effect of the collisions that we want to
constrain. 
 We find $M/L_B = 471 \pm 28, 422 \pm 25$ 
and $M/L_I = 179 \pm 11, 214 \pm 13$ for the
subcluster and the main cluster, respectively (for a discussion of the
errors on the mass measurements, see B06). The ratios agree with
one another to within about the 68\% confidence intervals.    
From the $I$ band data, we find that the ratio of $M/L$ ratios of the
subcluster and main cluster is $0.84 \pm 0.07$.  We conservatively
choose to use the $I$ band data only, since we want to put a firm
lower limit on this ratio, and $M/L_B$ is larger for the subcluster
than for the main cluster.
Assuming
each cluster
started out with similar $M/L$ values, which appears to be a
reasonable assumption for 
clusters in general (e.g., Mellier 1999; Dahle 2000), we
conclude that the subcluster could not have lost more than $\sim23\%$
of its initial mass. 
A comparison with the results from simulations
plotted in Figure~\ref{fig:sigma} (dashed line) shows that this implies
$\sigma/m \la 0.6$~cm$^2$~g$^{-1}$, which is a slight improvement over
the
previous best limit of $\sigma/m \la 1$~cm$^2$~g$^{-1}$ from the
conservative estimates of M04.

\subsection{Structure in Subcluster Dark Matter Distribution} \label{sec:substruct}

M04 suggested that scattered DM particles, which would account for about 1/5
of the total subcluster mass, might form tail features in
the DM distribution, similar to the tails seen in the X-ray image of
the gas bullet (see Figure~\ref{fig:image}).
The simulations allow us to determine whether the non-observation of
such tails in the mass map could be used to constrain $\sigma/m$.
We find that, rather than forming a tail, the scattered
particles are mostly deposited in the core of the main cluster, and 
do not form any features at a level
that is interesting for constraining $\sigma/m$.

\section{Discussion} \label{sec:discuss}

\subsection{Non-zero Impact Parameter} \label{sec:impact}

As M04 argue,
the morphology of the X-ray image, in particular, the symmetry of the
North-South X-ray bar (most likely an oblate spheroid viewed edge-on) between
the main cluster and subcluster mass
peaks around the axis of symmetry set by the shape of the X-ray bullet
(which gives its present velocity direction), combined with the
line-of-sight velocity and X-ray derived Mach 
number, indicate a merger axis that is $\sim10\arcdeg$ from the plane
of the sky, and that the cluster cores must have passed close to one
another, certainly within the $\sim 200$~kpc core radius of the main cluster.
In all simulations previously discussed, it was assumed that the bullet
subcluster passed directly through the center of the main cluster
core, i.e., that the impact parameter of the merger, $b$, is zero.
For $b > 0$, we expect that the effects of self-interacting DM will be
reduced, since the density is at a maximum when the core centers pass
directly through one another, and the scattering probability is
proportional to the density (Eqn.~\ref{eqn:prob}).  
To test the strength of this effect, we
re-ran the simulation R4 (see Table~\ref{tab:ics}) with an
impact parameter of $b = 200$~kpc.  Aside from the impact parameter,
the initial mass and velocity distributions were identical to those
for run R4, so that the relative
effects of $b >0$ could be investigated (specifically, no adjustments
were made to the initial conditions to more closely match the current
observed mass profiles).  The resulting projected total mass
profiles for the subcluster within 150~kpc and 250~kpc at the observed
separation agreed with those from the $b = 0$ run to within 4\%.  For the main
cluster, the match was better than 1\%.

The resulting offset between the galaxy and DM centroids during the
post core passage phase was
systematically smaller than the offset seen in the $b = 0$ run.  
At the observed separation, the difference in the centroid offsets, as
compared to the $b=0$ run, was
about 4~kpc, which
is on the order of both the observational error and the accuracy of our
numerical technique.  The fractional change in the $M/L$ ratio of the
subcluster was similarly affected; for the $b = 0$ case, the $M/L$ ratio
within 150~kpc drops by about 27\%, whereas for
the run with $b = 200$~kpc, it drops by 22\%.  Assuming, as we did in
\S~\ref{sec:mtol}, that the subcluster could not have lost more than
$\sim23$\% of its initial mass, we find the constraint that $\sigma/m <
0.7$~cm$^2$~g$^{-1}$.  We therefore conclude 
that, although a non-zero impact parameter reduces the effects of
self-interacting DM as expected, the level of the effect is
relatively small.  This is likely due to the assumed King mass profile of the
main cluster.  The radial density gradient is relatively small within
the core radius of the main cluster (which in this case is 151~kpc),
so it is not surprising that the effects of self-interacting DM are
not significantly reduced by increasing the impact parameter, so long
as it is comparable to the core radius of the main cluster.  Naturally,
the effects of a non-zero $b$ would be increased if the main cluster
had a strongly peaked mass profile, though the current lensing data
suggest otherwise.
We conclude that any value of impact parameter that is consistent with
the observations will only slightly alter our results.

\subsection{Alternative Bullet Mass Profiles} \label{sec:massprof}

As noted in \S\ref{sec:sims}, the choice of a King mass profile
for the subcluster is expected to give conservative estimates on the
effects of collisional DM, based on the $M/L$ ratio, since the central
density is low as compared 
to models with cuspy cores such as NFW and Hernquist models.  However,
in the case of the galaxy/DM centroid offset test, one might argue that
since the subcluster galaxies are more tightly bound in the center for
more highly concentrated mass profiles, it
will be more difficult to displace them, which could lead to a smaller
offset between the centroids despite the increased action of DM
collisions.  We therefore ran a test
simulation, using a King model for the main cluster and a Hernquist
model for the subcluster, with $\sigma/m = 0.72$~cm$^2$~g$^{-1}$.  
The Hernquist profile is given by
\begin{equation} \label{eq:hern}
\rho(r) = \frac{M a}{2 \pi r} \frac{1}{(r + a)^3},
\end{equation}
where $M$ is the total cluster mass and $a$ is the scale length
(Hernquist 1990).  As before, initial parameters were chosen such
that the bullet mass profile roughly matches the observed profile at
the current separation (we used $M = 3.13 \times 10^{14}$ M$_\odot$, $a
= 100$ kpc).  This is expected to be the most conservative model
combination for this test, since the main cluster King model minimizes
the effects of DM self-interaction while the subcluster Hernquist
model maximizes the binding energy of the subcluster galaxies.
The results show that, when comparing to run R4 in
Table~\ref{tab:obscond}, the galaxy/DM centroid offset was
only slightly less than that found with the King model subcluster, on
the order of the accuracy of the simulation offset values (less than 1~kpc).
The change in 
the subcluster's $M/L$ was similarly only weakly affected ($f =
0.27$ for the King model bullet subcluster, whereas for the Hernquist
model we find $f = 0.31$, consistent with the King profile being the
conservative case).  The agreement
is likely due to the fact that the centroids become offset from one
another after core passage, and it is during core passage that the
central density peak of the bullet is mostly ``smoothed away'' due to
DM collisions (recall that DM scattering is more frequent in high
density regions, so high density structures are more efficiently
destroyed by DM self-interactions).  
Although strongly peaked density profiles have been found to be
unstable to SIDM (e.g., Burkert 2000; Yoshida et al.\ 2000b),
in our simulation a significant change in density only occurred at
small radii, such that the total projected mass of the subcluster within
50~kpc remained stable up until the merger event.  Therefore the
subcluster mass distribution remained significantly more peaked
than a King profile cluster with the same projected mass within
150~kpc.  
We conclude that our results are
only weakly dependent on the mass profile chosen for the bullet, so
long as we require that the observed mass profile is reproduced.  For
the initially more centrally concentrated profiles, the effects of the
increased
binding energy in the core are balanced by the increased scattering
frequency in this region.

\subsection{Mass Profile Dependence} \label{sec:lowmass}

As mentioned in \S~\ref{sec:sims}, weak lensing is expected to
underestimate the mass of the lens by 10-20\%.  
There are two separate effects that contribute to this underestimation.
First, near the core of a cluster there is a large region without weak
lensing galaxies, and this region is effectively smoothed over when
computing the mass map.  Additionally, galaxies near the regions where
strong lensing dominates are measured in the weak lensing
approximation, which also leads to an underestimate of the mass in the
core.  Second, the total cluster mass can be underestimated due to
mass-sheet degeneracy, where the mass-map is affected by the
non-detection of mass at the edge of the field of view.
Although projection
of foreground and/or background structures unassociated with the
clusters will artificially increase the mass, it is highly unlikely
that such projected structures significantly contributed to the
detected lensing signal (C06).  Results from strong lensing, which is
not susceptible to the same systematic underestimation as weak
lensing, do indeed give systematically higher projected masses for
this system, by about a factor of 2 within the inner few hundred kpc,
which is the region we are most interested in for this analysis
(compare B06 and C06; see C06 for further discussion of this
discrepancy).  Though we
chose to use the mass estimates from strong lensing, since it should give a
more reliable estimate of the projected mass near the
cluster cores, it is interesting to explore the
dependence of our results on the lensing mass estimates.  To this
end, we conducted a simulation run similar to run R4, but with the
initial cluster central densities chosen such that the projected mass
profiles at the observed separation were about 2 times lower than the
masses derived from strong lensing observations, roughly in agreement
with the weak lensing results given by C06 (the initial
core radii were the same as for run R4).  Since the scattering
probability depends on the density, we expect these less
massive halos to be more weakly affected by SIDM.  

Results obtained
from the simulations 
with the lower mass normalization show that
the effects of SIDM are diminished, as expected for a linear
dependence of the scattering probabilities on the projected mass.
In run~R4, the $M/L$ ratio
dropped by 27\%, whereas for the run with 1/2 the total mass it
dropped by 14\% (roughly a factor of 2 less).  Similarly, the galaxy/DM
centroid offset was 11.1~kpc, again, about a factor of 2 down from the
24.1~kpc offset seen in run~R4.  If we assume that this factor of two
effect can be applied to all of values given in columns~6 \& 7 of
Table~\ref{tab:obscond} and consider the more sensitive $M/L$ test, we
find that requiring $f \la 0.20$ would correspond to $\sigma \la
1.25$~cm$^2$~g$^{-1}$.
This is done as a test of the method only, since these low halo mass
values are not realistic, as they are insufficient to produce the system
of strong arcs observed in the {\it HST} images (C06).

\subsection{Low Merger Velocity}\label{sec:lowvel}

 All of the simulations discussed so far have assumed a merger velocity
 that is consistent with that derived from X-ray observations
 (Markevitch, 2005),
 which give a Mach number for the shock front of $M = 3.0 \pm 0.4$,
 and it is assumed that the subcluster has the same velocity as the
 shock (though see Springel \& Farrar, 2007).  
 Since the subcluster could have slowed down, or the shock
 front accelerated, it is interesting to 
 ask what effect a lower velocity would have on the inferred upper limit on
 $\sigma/m$, particularly since the observed velocity is larger than
 would be expected from free fall of the subcluster onto the main
 cluster (Farrar \& Rosen 2007).
 In order to test the
 dependence of our upper limit on merger velocity, we ran a simulation
 with $\sigma/m = 0.72$~cm$^2$~g$^{-1}$ such that the relative velocity
 of the cluster DM halos at observed separation was 1.5 times lower,
 about 3100~km~s$^{-1}$ ($M \approx 2$).  This is close to the
 expected free-fall velocity of the subcluster, and to the relative
 velocity of 2860~km~s$^{-1}$ found by Springel \& Farrar (2007) from
 hydrodynamical simulations of this system.  The results showed
 little difference
 from the higher velocity run (compare to run R4 in 
 Table~\ref{tab:obscond}): $\Delta x$ was 30.2~kpc (vs. 24.1~kpc) and
 $f$ was 0.25 (vs. 0.27).  We therefore conclude that our results are
 relatively insensitive 
 to merger velocities that are not in large disagreement with the
 observations.  
 A weak dependence of the $M/L$
 ratio on the subcluster velocity $v$ is easy to understand: the
 particle scatters out of the subcluster as long as $v/2$ is much
 greater than the escape velocity from the subcluster, which it is by 
 a large margin (M04).

\subsection{Effects of Diffuse Gas}\label{sec:gas}

As mentioned in \S~\ref{sec:sims}, the intracluster gas observed in
the X-ray band was not included in the simulations (doing so would
greatly increase the computing time and the complexity involved with
matching the observations in detail).  The only way
for the gas to affect the results is via gravitational interaction (we
ignore the possibility of non-gravitational baryon-DM interactions,
the cross-section of which has been shown to be extremely small, e.g.,
Chen et al. 2002).  In general, the gas is expected to contribute
about 10\% of the total mass of the system, a figure which appears to
be consistent with the lensing and X-ray observations (B06).  
One might worry that, when matching the observed
mass profiles, some ``extra'' DM is needed to account for the missing
gas.  As mentioned in \S~\ref{sec:sims}, gas masses have been
subtracted from the lensing masses using a detailed model of the gas
distribution derived from fitting the X-ray observations.
In terms of the test involving the decrease in
the subcluster $M/L$ ratio (see
\S~\ref{sec:mtol}), we needn't worry about the subcluster gas for
the simple reason that the gas bullet is far from the subcluster mass
peak (roughly 23$\arcsec$, or 102~kpc).  Therefore,
the gas in the region of the bullet mass peak is not centrally
concentrated and will
not significantly add to the binding energy of potentially scattered
DM particles.  For the galaxy and total mass centroid offset test (see
\S~\ref{sec:noffset}), the exclusion of the gas is expected to give a
conservative result: the gas bullet and bar feature seen in
Figure~\ref{fig:image} will act to decelerate the subcluster DM halo
and galaxies.  However, if, as is the case with SIDM, the DM halo
starts to lag behind the galaxies and gets
closer to the gas cores than the main concentration of the galaxies, it
will experience a larger deceleration, thereby increasing the offset
between the two.  Due to the relatively low mass of the gas
components, and the large distance between the gas peaks and the subcluster
DM halo and galaxies (as compared to the offset of the latter), the
strength of this effect will be quite small.  We therefore conclude
that including gas in the simulations would not significantly affect
our results.

\section{Summary} \label{sec:summ}

We have combined results from new X-ray, optical and lensing
observations and our N-body simulations of the merging galaxy cluster
1E~0657-56 in order to derive an upper limit on the self-interaction
cross-section of dark matter particles, $\sigma/m$.  We give constraints on
$\sigma/m$ based on two independent methods: from the lack of offset between the
total mass peak and galaxy centroid of the subcluster that would arise during
the merger due to drag on the subcluster halo from DM particle
collisions, and from the lack of a decreased mass-to-light ratio of the
subcluster due to scattering of DM particles.  From
the former, we find $\sigma/m < 1.25$~cm$^2$~g$^{-1}$, and from the latter,
$\sigma/m < 0.7$~cm$^2$~g$^{-1}$, which includes the
uncertainty in the impact parameter of the merger (upper limits are
from 68\% confidence intervals).  
Our best constraint is a
modest improvement of the previous best constraint from conservative
analytic estimates
of $\sigma/m < 1$~cm$^2$~g$^{-1}$ (M04). 
Furthermore, our limit of $\sigma/m < 1.25$~cm$^2$~g$^{-1}$ is more
robust than the best analytic limit, since this method does not depend
on the assumption that the subcluster and main cluster $M/L$ ratios
were equal prior to the merger.
Previous studies have found that
$\sigma/m \sim 0.5 - 5$~cm$^2$~g$^{-1}$ is needed produce the
observational effects that self-interacting dark matter has been
invoked to explain (e.g., non-peaked galaxy mass profiles and the
underabundance of small halos within larger systems).  Our results rule
out almost this full range of values, at least under the assumption
that $\sigma$ is velocity-independent.

We would like to
thank Volker Springel, Naoki Yoshida, Yago Ascasibar, and Alexey
Vikhlinin for useful
discussions and for providing access to various private codes.
Simulations were performed on a Beowulf cluster at the ITC in the
Harvard-Smithsonian Center for Astrophysics.  Support for this work
was partially provided for by the NASA {\it Chandra} grants G04-5152X and
TM6-7010X, and NASA contract NAS8-39073.

\clearpage

\begin{deluxetable}{ccccccc}
\tablewidth{5.8truein}
\tablecaption{Initial Simulation Parameters \label{tab:ics}}
\tablehead{
\colhead{Run Name}&
\colhead{$N_{DM}$}&
\colhead{$\sigma/m$}&
\colhead{$\rho_{c,1}$}&
\colhead{$r_{c,1}$}&
\colhead{$\rho_{c,2}$}&
\colhead{$r_{c,2}$}\\
\colhead{}&
\colhead{}&
\colhead{(cm$^2$ g$^{-1}$)}&
\colhead{(10$^6$ M$_\odot$ kpc$^{-3}$)}&
\colhead{(kpc)}&
\colhead{(10$^6$ M$_\odot$ kpc$^{-3}$)}&
\colhead{(kpc)}
}
\startdata
R1 &10$^6$&0    &3.27 &213 &4.59  &149\\
R2 &10$^6$&0.24 &3.27 &213 &4.59  &149\\
R3 &10$^6$&0.48 &4.42 &183 &6.57  &129\\
R4 &10$^6$&0.72 &7.03 &151 &11.75 &108\\
R5 &10$^6$&0.96 &6.26 &167 &9.76  &124\\
R6 &10$^6$&1.25 &6.26 &167 &9.76  &124\\
\enddata
\end{deluxetable}

\clearpage

\begin{deluxetable}{ccccccc}
\tablewidth{6.7truein}
\tablecaption{Conditions at Observed Separation \label{tab:obscond}}
\tablehead{
\colhead{Run Name}&
\colhead{$N_{DM}$}&
\colhead{$\sigma/m$}&
\colhead{$M_1(r < 150 $kpc$)$}&
\colhead{$M_2(r < 150 $kpc$)$}&
\colhead{$\Delta$ x\tablenotemark{a}}&
\colhead{$f$\tablenotemark{b}}\\
\colhead{}&
\colhead{}&
\colhead{(cm$^2$ g$^{-1}$)}&
\colhead{(10$^{13}$ M$_\odot$)}&
\colhead{(10$^{13}$ M$_\odot$)}&
\colhead{(kpc)}&
\colhead{}
}
\startdata
R1  &$10^6$&0   &12.0& 11.1& 1.8& 0.0\\
R2  &$10^6$&0.24&11.5& 10.4& 5.4&0.08\\
R3  &$10^6$&0.48&11.8& 10.4& 15.0&0.16\\
R4  &$10^6$&0.72&12.6& 11.0&24.1&0.27\\
R5  &$10^6$&0.96&12.4& 10.9&37.9&0.32\\
R6  &$10^6$&1.25&11.4& 9.8&53.9&0.38\\
Obs.&      &    &11.9$\pm$1.6& 10.6$\pm$0.4&$25 \pm 29$ &$0.16\pm0.07$\\
\enddata
\tablenotetext{a}{$\Delta$ x is the offset between the subcluster
  total mass and galaxy centroids.}
\tablenotetext{b}{$f$ is the fractional decrease in the mass-to-light
  ratio of the subcluster within 150~kpc.}
\end{deluxetable}

\clearpage

\clearpage

\clearpage

\clearpage

\clearpage

\end{document}